# Cryogenic fiber-coupled electro-optic characterization platform for high-speed photodiodes

**Shekhar Priyadarshi**[(a)*], **Hao Tian**[(a)], **Alexander Fernandez Scarioni**[(a)], **Silke Wolter**[(a)], **Oliver Kieler**[(a)], **Johannes Kohlmann**[(a)], **Jaani Nissilä**[(b)], and **Mark Bieler**[(a)]

[(a] Physikalisch-Technische Bundesanstalt, Bundesallee 100,
38116 Braunschweig, Germany

[(b)]VTT Technical Research Centre of Finland Ltd,
02150 Espoo, Finland

[*]Corresponding author: shekhar.priyadarshi@ptb.de

We have developed a cryogenic characterization platform for ultrafast photodiodes, whose time domain responses are extracted by electro-optic sampling using femtosecond laser pulses in a pump-probe configuration. The excitation of the photodiodes with the pump beam and the electro-optic sampling crystals with the probe beam are realized in a fully fiber-coupled manner. This allows us to place the characterization platform in almost any temperature environment. As application example, we characterize the time-domain response of commercial p-i-n photodiodes with a nominal bandwidth of 20 GHz and 60 GHz at temperatures of 4 K and 300 K and in a large parameter range of photocurrent and reverse bias. For these photodiodes, we detect frequency components up to approximately 250 GHz, while the theoretical bandwidth of our sampling method exceeds 1 THz. Our measurements demonstrate a significant excitation power and temperature dependence of the photodiodes' ultrafast time responses, reflecting, most likely, changes in carrier mobilities and electric field screening. Since our system is an ideal tool to characterize and optimize the response of fast photodiodes at cryogenic temperatures, it has direct impact on applications in superconducting quantum technology such as the enhancement of optical links to superconducting qubits and quantum-accurate waveform generators.

**Introduction**

Ultrafast photodiodes are important components for a variety of applications such as photonic integrated circuits and high-speed optical communication and sensing[1,2], which are typically operated in room-temperature environments. Yet, with the advancement of superconducting quantum technology, additional application scenarios in other temperature ranges evolved for ultrafast photodiodes. These scenarios are, for example, optical links from room-temperature electronics to superconducting qubits at mK temperatures[3] and optical-pulse-driven quantum-accurate waveform generators at a temperature of 4 K[4]. The main drive for the development of optical links in combination with cryogenic photodiodes are the reduced heat insertion into the cryogenic environment[5] and reduced high-frequency signal distortion as compared to coaxial cables.

For basically all applications involving high-speed photodiodes it is essential to know their time- or frequency-domain responses. So far, corresponding room-temperature characterization techniques have been developed and enhanced over more than a decade[6–9]. National Metrology Institutes even performed an international comparison on the room-temperature, time-domain response of a photodiode with a nominal bandwidth of 100 GHz[10]. With respect to low-temperature environments, cryogenic measurements of time-domain waveforms have already been performed for certain applications[11–13]. However, accurate characterization platforms for high-speed photodiodes operating at cryogenic temperatures do not exist. Although this is very important, since the scattering rates and mobilities of charge carriers changes considerably when going from room-temperature to the cryogenic environment. As a consequence, the time- and frequency domain responses might be very different at cryogenic temperature as compared to room temperature.

In this paper, we present a fully fiber-coupled characterization platform operating at cryogenic temperatures. In this platform, voltage pulses are generated by pulsed-laser excitation of the photodiode under test, travel subsequently on a coplanar waveguide (CPW), and are finally detected employing electro-optic (EO) sampling with another pulsed laser. The photodiodes under test are commercially available p-i-n photodiodes with nominal bandwidths of 20 GHz and 60 GHz. The EO sampling is realized in a lithium tantalate (LTA) crystal that was attached to the CPW. The main advantage of our setup is that it can be placed in almost any temperature environment and still maintains the capability of ultrashort time and ultrahigh frequency resolution. In our experimental setup we measured spectral components generated by the photodiodes up to 250 GHz, with the theoretical bandwidth of the EO detector exceeding 1 THz[12,14]. We study the time-domain response of the photodiodes in a large parameter range at temperatures of 4 K and 300 K, focusing on the transition between the linear and the saturation regime. We observe clear changes of the photodiodes' ultrafast response when varying the excitation power and the temperature. The different carrier dynamics are most likely linked to different mobilities of electrons and holes in combination with field screening effects. These results directly demonstrate the usefulness of our characterization platform for optimizing photodiode responses, in particular, for applications with superconducting circuits[4,15–18].

**Material and Methods**

The main part of the characterization platform is shown in Fig. 1(a). The fabrication starts with 380-µm thick high-resistivity (20 kΩ cm) silicon (Si) wafers with a 300 nm thick thermally-grown $SiO_2$ insulation layer on top. On these wafers, resist masks for the CPW structures were defined employing a standard electron-beam-lithography process. The metallization forming the CPW was carried out by subsequent sputtering of a 5 nm titanium (Ti) layer and a 500 nm gold (Au) layer without interrupting the vacuum. The Ti layer serves as an adhesion layer for the Au. The unwanted metallization was removed after the sputter deposition by a lift-off process in acetone. The CPWs fabricated in this way have the following dimensions: 8.5 mm long, 29 µm width of the signal conductor, 200 µm width of the ground conductors, and 18 µm gap between signal and ground conductors. These dimensions result in a nominal 50 Ω impedance of the CPW in a large frequency range.

As PDs under test, we used two different types of commercially available p-i-n PDs with InGaAs as the intrinsic region and a nominal bandwidth of 20 GHz and 60 GHz. The PDs have to be backside-illuminated through an integrated lens, which simplifies the alignment of the optical excitation beam[4]. The PDs are mounted on the left-hand-side end of the CPW, see Fig. 1(a) using flip-chip bonding technology. To this end Au stud bumps were placed on the contact pads forming the end of the CPW. The PDs were then placed on the stud bumps and bonded in a flip-chip machine using thermo-compression. During bonding a force of 4 N - 5 N was applied at a temperature of 250°C. The terminals at the right-hand-side end of the CPW were wire bonded to a printed circuit board (PCB) with the Si substrate being glued to the PCB. The PDs were reverse biased using the electrical connections on the right-hand-side of the CPW and an externally connected bias tee, which allowed 50 Ω termination of the CPW for frequencies larger than 200 kHz. The PDs are optically excited using approximately 200 fs (full width at half maximum, FWHM) long pulses with a center wavelength of 1340 nm and a repetition rate of 76 MHz. (We explain the realization of the fiber-chip coupling further below.) The choice of 1340 nm wavelength as pump was important, since excitation of the PDs with 1550 nm wavelength hardly yielded any photocurrent at 4 K due to the temperature dependence of the bandgap of InGaAs. The total length of the SMF-28 pump fiber was approximately 6.5 m long and due to low chromatic dispersion around 1340 nm, the pump pulses maintained their temporal width during propagation through these fibers.

The voltage pulses generated by the PDs, subsequently propagate along the CPW. To detect the voltage pulses, we placed an LTA crystal on top of the CPW. The LTA crystal was 50 µm thick and X-cut with crystallographic Y and Z axes being aligned parallel and perpendicular to the CPW, respectively[19]. The LTA crystal had an anti-reflection coating on the top surface on which the probe pulses were incident, and a high reflection (HR) coating on the bottom side, which was placed on the CPW. (As for the pump fiber, we explain the fiber-chip coupling for the probe fiber further below.) The probe pulses also have a pulse width of approximately 200 fs and a repetition rate of 76 MHz. However, in contrast to the pump pulses the optical probe pulses with 1550 nm center wavelength experience strong dispersion of approximately 18 ps/(nm km) in SMF-28 fibers. This dispersion broadens ~ 200 fs long probe pulses to ~ 3 ps after a propagation distance of 1.5 m in the fiber. To achieve high time resolution it was very important to obtain short probe pulses at the sampling position. Accordingly, we used a combination of a dispersion-compensating fiber, which induced negative dispersion, and an SMF-28 fiber (the length of both fibers together was approximately 6 meters) to realize a probe pulse FWHM of approximately 200 fs at the sampling position.

In addition to the FWHM of the probe pulse, two additional time constants define the time resolution of our experimental setup. The first time constant is given by the time that the voltage pulse needs to propagate through the focal region of the probe spot (~10 µm). In turn, the second time constant is given by the time that the probe pulse needs to propagate through the electric field region of the voltage pulse. Since the electric field extends completely into the 50 µm thick LTA crystal, the second time constant equals the propagation time of the probe pulse through the LTA crystal. Simple calculations show that the second time constant is significantly larger than the first time constant and the FWHM of the probe pulse leading to an overall time resolution of approximately 0.7 ps. This time constant is well suited to detect frequency components even above 1 THz.

The electric field of the voltage pulses induced an additional birefringence in the LTA crystal due to electro-optic effect. Consequently, the probe pulses experience an electric-field induced polarization change when propagating through the LTA crystal. The HR coating of the LTA crystal led to a back reflection of the probe pulses into the fiber, in which the back reflected pulses were separated from the incident pulses using a fiber circulator. Subsequently, the electric-field-induced polarization change of the reflected probe beam was analyzed employing ellipsometric techniques[20] with the resulting signal being proportional to the voltage value at the particular sampling instant. To map the temporal shape of the voltage pulses we varied the time delay between the pump and probe pulses. For this purpose we synchronized the repetition rate of the two lasers (fundamental and 15$^{th}$ harmonic) to each other, with the pump and probe lasers being the master and slave, respectively. With this synchronization scheme, we can easily and accurately vary the time delay between pump and probe pulses by adding an offset to the phase of the 15$^{th}$ harmonic signal. We then measured the voltage versus temporal delay between pump and probe employing a standard lock-in technique to increase the signal-to-noise ratio. We note that this sampling technique is different than conventional asynchronous sampling techniques, where lock-in detection is typically not employed[21].

To perform experiments at cryogenic temperatures, we mounted the PCB as shown in Fig. 1(a) on a custom designed sample rod, which we inserted into a liquid Helium Dewar. To deliver pump and probe pulses to the Si chip, we used the optical fibers as described above. In the vicinity of the PD and the EO crystal, the fibers ended in a borosilicate ferrule. One of the most difficult tasks related to this work was to align and attach these ferrules to the PD and the EO crystal in a robust manner and these processes are described in the following. To align the pump pulses to the PD, we first inserted the ferrule into a borosilicate sleeve. We then adjusted the position and angle of the sleeve with respect to the Si chip by maximizing the photocurrent generated in the PD by the pump beam. Once the photocurrent was maximized, we applied a commercially available two-component epoxy adhesive along the outer edges of the bottom of the sleeve. Pressing the sleeve against the chip with a relatively small pressure of several N, we allowed the adhesive to cure for at least 15 hours. During the application of adhesive, we ensured that no adhesive crept between the ferrule and PD. We have investigated different two-component epoxy adhesives and finally chosen the product UHU plus endfest for all gluing processes as it survived the largest number of cooldown cycles and was easiest to handle. We note that other adhesives might perform equal or even better. Alternative ways for realizing fiber-PD coupling are described in references [4,18].

For the alignment of the probe fiber we proceeded in a slightly different way. First the LTA crystal was glued to the Si substrate. We did not glue the probe sleeve directly to the LTA crystal, since its thickness of 50 µm makes it very fragile and difficult to handle. Instead, we mounted a 1 mm thick sapphire crystal, which had a hole with a diameter equal to the outer diameter of the sleeve, on top of the LTA crystal, see Fig. 1(a). Here the sapphire crystal extends beyond the LTA crystal on the two sides perpendicular to the CPW, see Fig. 1(a) and glue was applied to these extensions, thus, realizing a direct adhesion of the Si substrate and the sapphire crystal. Before the glue was cured, we inserted the probe beam sleeve into the hole of the Sapphire slab and positioned it to obtain a maximum EO signal ($EO_{\text{ref}}$). To this end, the $EO_{\text{ref}}$ was obtained with the pump beam being blocked and no reverse bias being applied to the PD. Instead, we applied a square signal with a frequency of 1 MHz and a unipolar amplitude of 1 V to the CPW. The electric-field-induced birefringence of the probe beam was read using the lock-in detection and the resulting signal corresponds to the $EO_{\text{ref}}$. After aligning the probe beam sleeve by maximizing the $EO_{\text{ref}}$, we applied additional adhesive at the juncture of the sleeve and the sapphire slab and allowed the whole adhesive to cure for at least 15 hours. During the application of adhesives, we ensured that no adhesive reached the bottom of the probe ferrule. We note that during gluing of the LTA crystal on the Si substrate a laminar film of glue is formed at the interface, lifting up the LTA crystal by a few µm. The advantage of the lifted LTA crystal is that it reduces the invasiveness of EO sampling[22]. The preparation of one sample as shown in Fig. 1(a) took approximately 3 working days.

**Results and Discussions**

We start describing experimental results by demonstrating the proper operation of the EO sampling method. We generated voltage pulses by excitation of the 20 GHz PD and detected the pulses at different probe positions perpendicular to the CPW. These measurements were done at room temperature and the whole arrangement for EO sampling consisting of LTA crystal, sapphire crystal and probe sleeve/ferrule was not glued to the CPW but could be moved perpendicular to the CPW. (Note that all other experimental results were obtained with glued probe-fiber-chip arrangement.) The reverse bias applied to the PD was $U_{\text{rev}} = 2$ V and the pump power was adjusted to result in an average photocurrent of $I_{\text{ph}} = 20$ µA. In general, we have chosen the operational conditions in this work such that a voltage pulse with a maximal amplitude of several 100 mV is generated. Thereby, we focus especially on the transition between the linear and the saturation regime of the PD. In Fig. 1(b), the voltage pulse maxima (expressed in lock-in units, black squares) and the $EO_{\text{ref}}$ (expressed in lock-in units, red circles) are plotted for different probe positions perpendicular to the CPW. The yellow background in Fig. 1(b) represents the metallization of the CPW. The evolution of the two signals is antisymmetric with a zero crossing at the center of the CPW and peaks in the two gaps between the signal and ground lines of the CPW. Figure. 1(c) shows time-resolved voltage pulses corresponding to the two peak values (red and blue) and the close-to-zero value (green) of the voltage pulse amplitudes plotted in Fig. 1(b). The voltage pulses corresponding to the two peaks of Fig. 1(b) have a similar temporal behavior with opposite polarity, and the voltage pulse corresponding to the center position on the CPW is very close to zero at all times, as expected. Moreover, the amplitudes of the voltage pulses and $EO_{ref}$ plotted in Fig. 1(b) have a nearly identical dependence on sampling position. These results increase the confidence in our measurements since the experimental setup is expected to be sensitive to electric field components of the voltage pulses being parallel to the Z axis of the LTA crystal. In following, we use $EO_{ref}$ to normalize the time-dependent voltage pulse measurements, thus, obtaining voltage pulses corresponding to the unit Volt.

We now address the time-and frequency-domain characterization of the PDs. In Figs. 2(a) and (b) are plotted the time traces of the voltage pulses measured at 4 K (solid blue) and 300 K (dashed red), obtained from the 20 GHz and 60 GHz PDs, respectively. During the measurements $U_{rev}$ and $I_{ph}$ were kept at 2 V and 20 µA, respectively, for the 20 GHz PD and at 1.2 V and 5 µA, respectively, for the 60 GHz PD. We observe that both, the temporal width and the tail of the voltage pulses decrease when reducing the temperature from 300 K to 4 K and when exchanging the 20 GHz PD with the 60 GHz PD. While the faster response for the 60 GHz PD is expected, the faster response and, in particular, the reduced tail at 4 K as compared to 300 K is unexpected. Yet, this reduced tail is advantageous for applications in superconducting technology since it corresponds to an improvement of PD operation at cryogenic temperatures. In Figs. 2(c) and (d), the spectra corresponding to the unnormalized time traces of Figs. 2(a) and (b) are shown. All spectra extend up to ~ 250 GHz before reaching the noise level at about -120 dB and the reduction of the tail in the time-domain goes along with increased spectral components at high frequencies as compared to very low frequencies.

In addition to the PD response also parameters of the measurement setup depend on the temperature. These parameters mainly influence the signal-to-noise ratio of the measurement, but not the time-domain shape of the PD response. Still, it is instructive to comment on these dependencies as well. The $EO_{ref}$ at 4 K corresponds to ~ 2/3 of the $EO_{ref}$ at 300 K. One reason for this change are the different optical[23] and electro-optical properties of the LTA crystal at the two temperatures. A certain change in $EO_{\text{ref}}$ might also result from a movement of the glued probe fiber with respect to CPW, which occurred during the cooldown due to different expansion coefficients of materials involved in the fiber chip coupling. The effect of such a movement, which is estimated to be on the order of a few µm, can be minimized by placing the probe fiber ferrule in a region where small spatial shifts do not lead to a significant change of $EO_{\text{ref}}$, i.e., close to the inner edges of the ground lines of the CPW, see Fig. 1b. In addition to $EO_{\text{ref}}$ also the photocurrent of the studied PDs changed by a factor of ~ 2/3 after temperature change from 300 K to 4 K. We believe that the main reason for the photocurrent decrease is the temperature dependence of the bandgap of the PD's absorption region, i.e., InGaAs. We note that movements of the pump fiber on the order of a few µm can be ignored due to the built-in lens on the backside of the PD[4]. For a comparison of measurements performed at 4 K and 300 K we kept the photocurrent constant by increasing the optical power of the pump beam at cryogenic conditions.

For the characterization of time-domain waveforms, the amplitude and temporal width of voltage pulses represent the two most important parameters. In Figs. 3 (a) and (b) are plotted the FWHM of the voltage pulses for the 20 GHz and 60 GHz PDs, respectively, for different $I_{\text{ph}}$ (blue squares) and $U_{\text{rev}}$ (red circles). The corresponding amplitude, i.e., peak, of the voltage pulses are plotted in Figs. 3 (c) and (d). The FWHM increases monotonically with increasing $I_{\text{ph}}$, resulting in a super-linear and linear behavior for the 20 GHz and 60 GHz PDs, respectively. The dependence of the FWHM on $U_{\text{rev}}$ is different, though. A minimum with 16.5 ps is obtained at $U_{\text{rev}} = 4$ V for the 20 GHz PD, whereas the FWHM of the 60 GHz PD decreases monotonically with increasing $U_{\text{rev}}$ and reaches a value of 10.5 ps at $U_{\text{rev}} = 1.2$ V. The amplitude of the voltage pulses increases sub-linearly with $I_{\text{ph}}$ for both PDs, showing that the saturation regime is reached for the large $I_{\text{ph}}$. Again, the dependence on $U_{\text{rev}}$ is different. A maximum is obtained for the 20 GHz PD, whereas a linear increase versus increasing $U_{\text{rev}}$ exists for the 60 GHz PD. The dependences of the FWHM and amplitude on $U_{\text{rev}}$ are approximately inverted to each other. This reflects the fact that the time-domain integral of the voltage pulse (which is a measure of the power produced by the PD) does not depend on $U_{\text{rev}}$. It should be emphasized that other dependencies might be obtained for different operational conditions. The results shown in Fig. 3 are important whenever the temporal width and amplitude of the voltage pulses need to be optimized.

Finally, we comment on the dependences of the ultrafast time-domain responses on the operating conditions, i.e., on $I_{\text{ph}}$ and $U_{\text{rev}}$. The non-monotonic behaviors shown in Fig. 3 already indicate complex current dynamics, which also becomes visible in the time-domain waveforms. In Figs. 4(a) and (c) voltage pulses obtained from the 20 GHz PD are plotted for different $U_{rev}$ and $I_{ph}$, respectively. Figures 4(b) and (d) show similar plots for the 60 GHz PD. In general, we observe two contributions to the time-domain response. The first (fast) contribution leads to the initial voltage pulse with a very fast rise time. The second (slow) contribution leads under certain operational conditions to a second time-delayed voltage peak and seems to be responsible for the enhanced tails of the voltage pulses. The dependence of the tail on the operational conditions becomes most prominently visible for the 20 GHz PD, whereas the second peak is most pronounced for the 60 GHz PD, but to a lesser extend also visible in Fig. 4(a) for the 20 GHz PD and $U_{rev}$ = 10 V. We note that in time-domain measurements performed at room temperature no clear second peak can be observed, see Figs. 1(c), 2(a), and 2(b). The observations suggest that the first main peak results from the electron contribution to the voltage response, whereas the second peak and the slowly decaying tail denote hole contributions[8,24]. In reverse-biased PDs, the photocurrent ceases as soon as all photocarriers leave the depletion region, which, in a simple picture, can be considered as an effective recombination process. For holes this effective recombination time is longer compared to electrons since holes have a smaller mobility. The ultrafast femtosecond excitation additionally complicates the current dynamics, since carrier generation and, thus, subsequent screening of the bias field happen on a timescale of several 100 fs. Moreover, in addition to the carrier dynamics in the depletion region also the transport and, thus, mobility of carriers in non-depletion regions has to be taken into account to fully model the ultrafast response of p-i-n PDs[25]. In any case, the observation of complex ultrafast responses of PDs directly demonstrates the usefulness of our characterization platform as corresponding measurements would have been impossible with conventional room temperature electronics.

**Conclusions**

In conclusion, we have presented a fully-fiber-coupled platform for in-situ voltage pulse measurements at cryogenic conditions. The measurement principle is based on opto-electronic generation and electro-optic detection of voltage pulses travelling on coplanar waveguides. To demonstrate the functionality of our setup at cryogenic temperatures, we characterized the time- and frequency-domain responses of commercially available photodiodes with nominal bandwidths of 20 GHz and 60 GHz. We are confident that our characterization platform will prove to be a valuable tool for further optimization and enhancements of cryogenic high-speed photodiodes that may be used to deliver high-frequency signals to superconducting quantum circuits or other cryogenic devices.

**Acknowledgements:** The authors thank Peter Hinze and Thomas Weimann for excellent technical support and acknowledge funding from the EMPIR programme co-financed by the Participating States and from the European Union's Horizon 2020 research and innovation programme under grant agreement 20FUN07 and from the European Union's Horizon 2020 research and innovation programme under grant agreement No 899558.

**Author contributions**: MB and JN initiated the study. SP, HT, AFS, SW, and OK fabricated the characterization platform, incl. fibre-chip coupling. Optical and electrical measurements and data analysis was carried out by SP. All authors discussed the results. SP and MB wrote the paper, which was reviewed by all authors.

**Competing Interests:** The authors declare that they have no competing interests.

**Data availability:** The data that support the findings of this study and details of the PDs such as manufacturer and product number can be obtained from the corresponding author upon reasonable request.

**References:**

1. Umezawa, T. *et al.* High speed photoreceiver over 100 GHz and its application. *2016 IEEE Photonics Conf.* 562–563 (2017).

2. Rouvalis, E. *et al.* 170 GHz Photodiodes for InP-based photonic integrated circuits. *Opt. Express* **20**, 20090 (2012).

3. Lecocq, F. *et al.* Control and readout of a superconducting qubit using a photonic link. *Nature* **591**, 575–579 (2021).

4. Bardalen, E. *et al.* Packaging and Demonstration of Optical-Fiber-Coupled Photodiode Array for Operation at 4 K. *IEEE Trans. Components, Packag. Manuf. Technol.* **7**, 1395–1401 (2017).

5. Youssefi, A. *et al.* A cryogenic electro-optic interconnect for superconducting devices. *Nat. Electron.* **4**, 326–332 (2021).

6. Struszewski, P., Pierz, K. & Bieler, M. Time-Domain Characterization of High-Speed Photodetectors. *J. Infrared, Millimeter, Terahertz Waves* **38**, 1416–1431 (2017).

7. Lee, D.-J. & Whitaker, J. F. An optical-fiber-scale electro-optic probe for minimally invasive high-frequency field sensing. *Opt. Express* **16**, 21587 (2008).

8. Ishibashi, T. & Ito, H. Uni-traveling-carrier photodiodes. *J. Appl. Phys.* **127**, 031101 (2020).

9. Williams, D. F. *et al.* Covariance-based uncertainty analysis of the NIST electrooptic sampling system. *IEEE Trans. Microw. Theory Tech.* **54**, 481–491 (2006).

10. Bieler, M. *et al.* International comparison on ultrafast waveform metrology. in *2020 Conference on Precision Electromagnetic Measurements (CPEM), IEEE* 414–415 (2020).

11. Griebel, M. *et al.* Picosecond sampling with fiber-illuminated ErAs:GaAs photoconductive switches in a strong magnetic field and a cryogenic environment. *Appl. Phys. Lett.* **82**, 3179–3181 (2003).

12. Gallagher, W. J. *et al.* Subpicosecond optoelectronic study of resistive and superconductive transmission lines. *Appl. Phys. Lett.* **50**, 350–352 (1987).

13. Wang, C. C. *et al.* Optoelectronic generation and detection of single-flux-quantum

pulses. *Appl. Phys. Lett.* **66**, 3325 (1995).

14. Krökel, D., Grischkowsky, D. & Ketchen, M. B. Subpicosecond electrical pulse generation using photoconductive switches with long carrier lifetimes. *Appl. Phys. Lett.* **54**, 1046–1047 (1989).

15. Kieler, O. *et al.* Optical Pulse-Drive for the Pulse-Driven AC Josephson Voltage Standard. *IEEE Trans. Appl. Supercond.* **29**, 1200205 (2019).

16. Kieler, O. *et al.* Stacked Josephson Junction Arrays for the Pulse-Driven AC Josephson Voltage Standard. *IEEE Trans. Appl. Supercond.* **31**, 1–5 (2021).

17. Nissilä, J. *et al.* Driving a low critical current Josephson junction array with a mode-locked laser. *Appl. Phys. Lett.* **119**, 032601 (2021).

18. Bardalen, E., Karlsen, B., Malmbekk, H., Akram, M. N. & Ohlckers, P. Evaluation of InGaAs/InP photodiode for high-speed operation at 4 K. *Int. J. Metrol. Qual. Eng.* **9**, 13–17 (2018).

19. Casson, J. L. *et al.* Electro-optic coefficients of lithium tantalate at near-infrared wavelengths. *J. Opt. Soc. Am. B* **21**, 1948 (2004).

20. Gallot, G. & Grischkowsky, D. Electro-optic detection of terahertz radiation. *J. Opt. Soc. Am. B* **16**, 1204 (1999).

21. Struszewski, P. & Bieler, M. Asynchronous Optical Sampling for Laser-Based Vector Network Analysis on Coplanar Waveguides. *IEEE Trans. Instrum. Meas.* **68**, 2295–2302 (2019).

22. Seitz, S. *et al.* Characterization of an external electro-optic sampling probe: Influence of probe height on distortion of measured voltage pulses. *J. Appl. Phys.* **100**, 113124 (2006).

23. Jacob, M. V. *et al.* Temperature Dependence of Permittivity and Loss Tangent of Lithium Tantalate at Microwave Frequencies. *IEEE Trans. Microw. Theory Tech.* **52**, 536–541 (2004).

24. Matavulj, P. S., Gvozdic, D. M. & Radunovic, J. B. Analysis of the linear and nonlinear time response of a P-i-N photodiode by a two-valley model. *Proc. Int. Conf. Microelectron.* **1**, 331–334 (1997).

25. Goushcha, A. O. & Tabbert, B. On response time of semiconductor photodiodes. *Opt. Eng.* **56**, 097101 (2017).

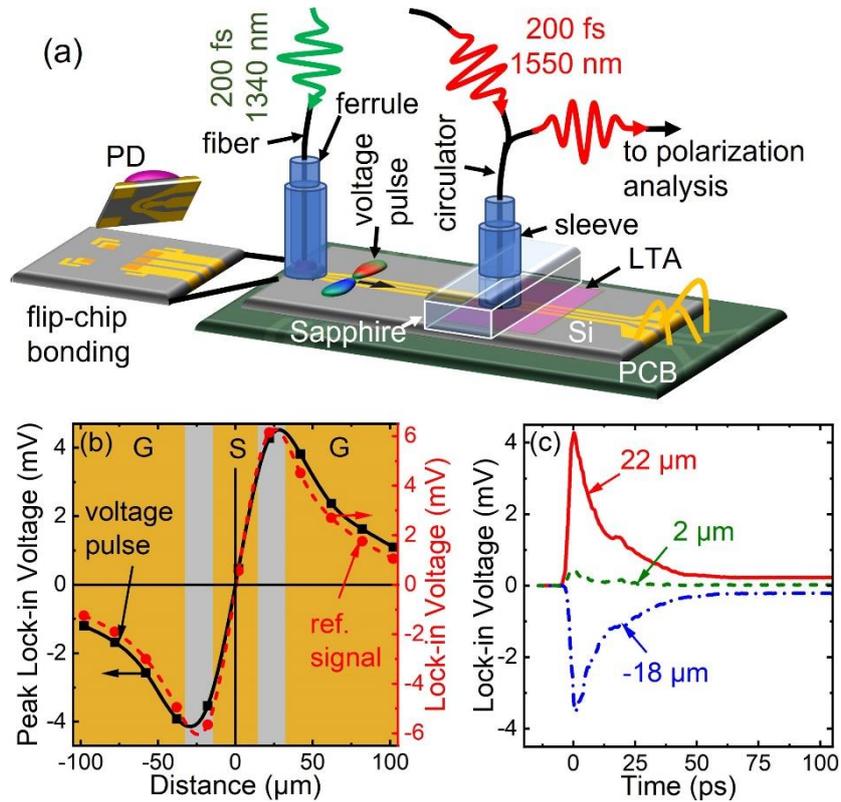

**Fig. 1.** (a) Electro-optic (EO) sampling scheme: The Photodiode (PD) is flip-chip mounted to the left-hand side of a coplanar waveguide (CPW), which is fabricated on a Si substrate. The Si substrate is mounted on a printed circuit board (PCB) and the CPW is connected to it via wire bonding. The coupling of the pump and probe fibers to the PD and the lithium tantalate (LTA) crystal, respectively, is realized using sleeves and ferrules. To protect the thin LTA crystal from breaking, a sapphire crystal is placed directly above the LTA crystal. The probe beam being reflected from the bottom of the LTA crystal is collected using a fiber circulator and guided to polarization analysis to extract an EO signal. The LTA crystal is X-cut with the crystallographic Y and Z axes being aligned parallel and perpendicular to the CPW, respectively. (b) Peak of the lock-in signal of a voltage pulse measurement (black squares) and lock-in signal resulting from the reference signal (red circles) versus distance perpendicular to the CPW. The colored background of the plot shows ground (G) and signal (S) conductors of the CPW. (c) Time-resolved lock-in voltages corresponding to the sampling positions -18 µm, 2 µm, and 22 µm perpendicular to the CPW with the center of the signal line being located at 0 µm.

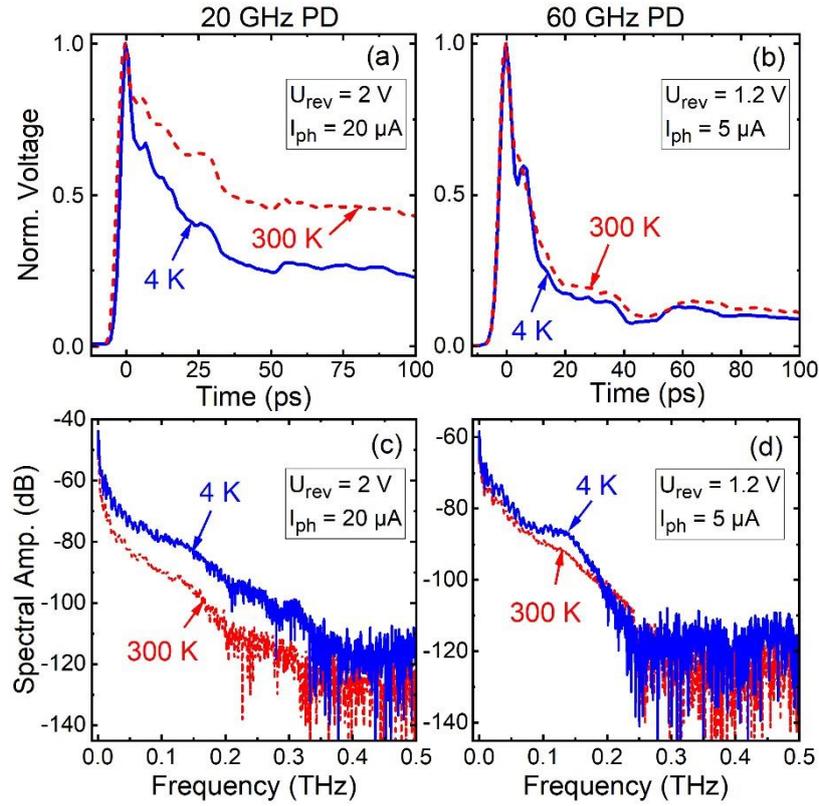

**Fig. 2.** Time traces of normalized voltage pulses obtained from a 20 GHz PD (a) and 60 GHz PD (b) at 300 K (red dashed line) and 4 K (blue solid line). The corresponding spectra of the unnormalized voltage pulse are shown in (c) and (d) for the 20 GHz PD and 60 GHz PD, respectively. The boxes in the figures denote the reverse bias $U_{\text{rev}}$ and photocurrent $I_{\text{ph}}$ for the different measurements.

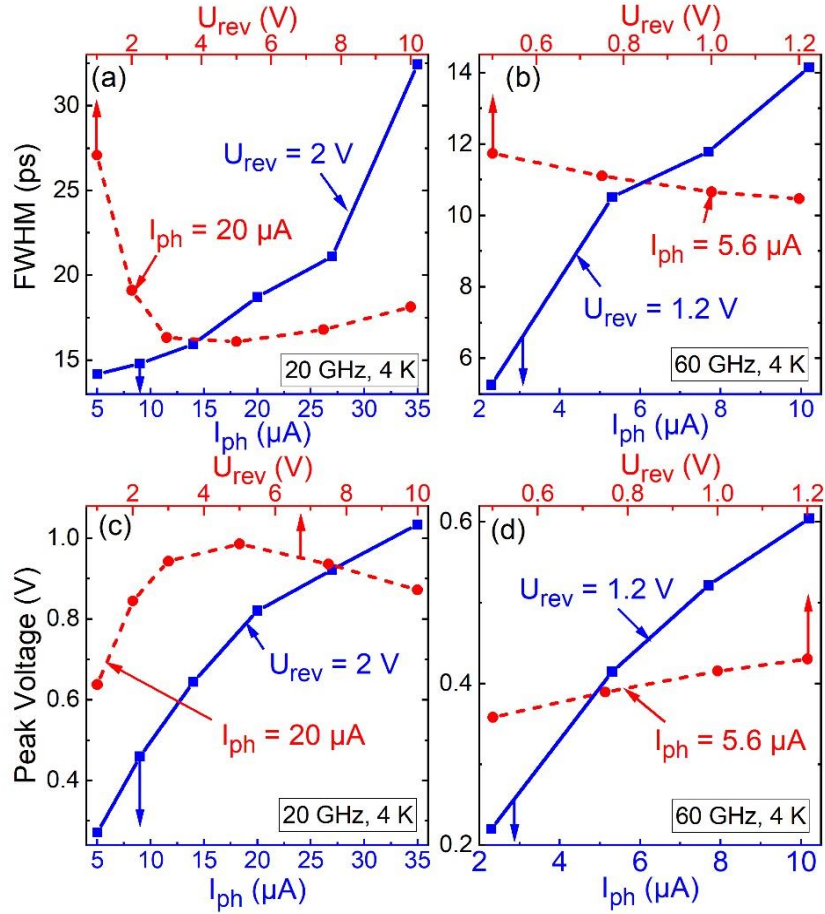

**Fig. 3.** Dependences of the full-width-at-half-maxima (FWHM) of the voltage pulses on the reverse bias $U_{\text{rev}}$ (red dashed line) and photocurrent $I_{\text{ph}}$ (blue solid line) for the 20 GHz (a) and 60 GHz (b) PDs. Dependences of the peak amplitude of the voltage pulses on the reverse bias $U_{\text{rev}}$ (red dashed line) and photocurrent $I_{\text{ph}}$ (blue solid line) for the 20 GHz (c) and 60 GHz (d) PDs. All measurements were done at a temperature of 4 K. The $I_{\text{ph}}$ settings corresponding to the $U_{\text{rev}}$ variation and, likewise, the $U_{\text{rev}}$ settings corresponding to the $I_{\text{ph}}$ variation are denoted in the figures.

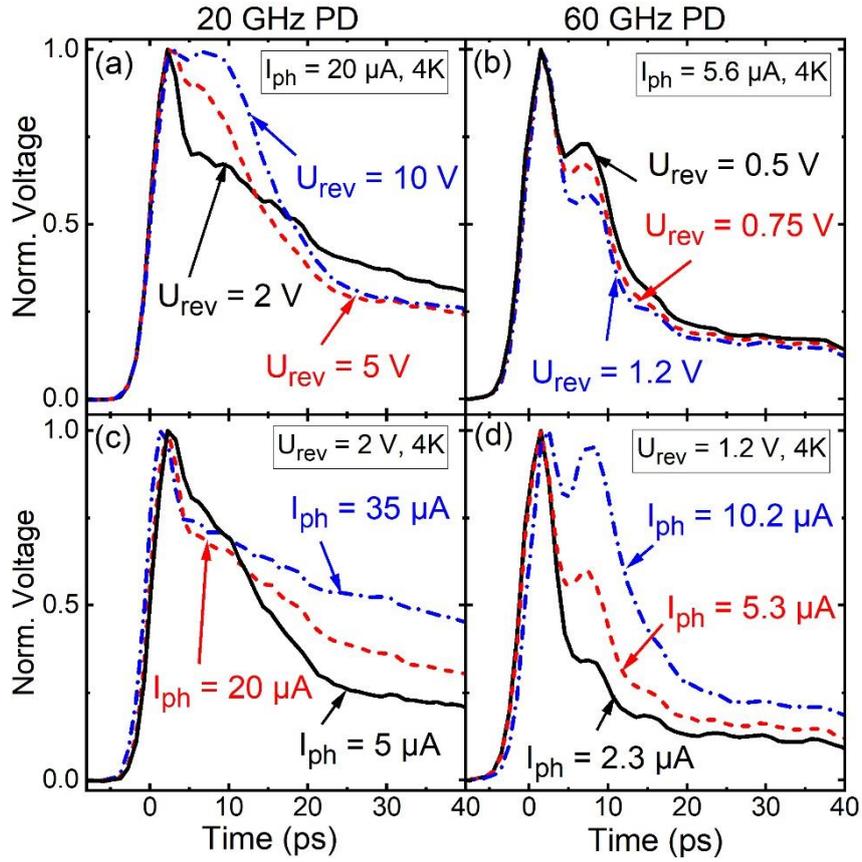

**Fig. 4.** (a) and (b) show time traces of normalized voltage pulses obtained for different reverse biases $U_{\mathrm{rev}}$ for the 20 GHz and 60 GHz PDs, respectively. (c) and (d) show time traces of normalized voltage pulses obtained for different photocurrents $I_{\mathrm{ph}}$ for the 20 GHz and 60 GHz PDs, respectively. All measurements were done at a temperature of 4 K. The $I_{\mathrm{ph}}$ settings corresponding to the $U_{\mathrm{rev}}$ variation and, likewise, the $U_{\mathrm{rev}}$ settings corresponding to the $I_{\mathrm{ph}}$ variation are denoted in the figures.